\documentclass[amsmath,amssymb,nofootinbib]{revtex4}
\pdfoutput=1

\usepackage{graphicx}
\usepackage{epsfig}
\usepackage{rotate}
\usepackage{amsmath}
\usepackage{amssymb}
\usepackage{amsfonts}
\usepackage{enumerate}
\usepackage{afterpage}
\usepackage{color}
\usepackage{amscd}
\usepackage{multirow}
\usepackage{xcolor}

\usepackage{amsfonts}
\usepackage{amsmath}
\usepackage{fancyhdr}
\usepackage{layout}
\usepackage{times}
\usepackage{natbib}
\usepackage{soul}
\usepackage[normalem]{ulem} 
\newif\ifAMStwofonts
\AMStwofontstrue

\newcommand{\fnl}{f_{\rm NL}}

\newcommand{\R}{{\zeta}}

\def\be{\begin{equation}}
\def\ee{\end{equation}}
\def\bea{\begin{eqnarray}}
\def\eea{\end{eqnarray}}

\begin{document}
\title{\vspace{-0.5cm}{\normalsize \normalfont \flushright YITP-15-47\\}
\vspace{0.6cm} A relativistic signature in large-scale structure}

\author{Nicola Bartolo$^{a,b}$, Daniele Bertacca$^{c,d}$, Marco Bruni$^{e}$, 
Kazuya Koyama$^{e}$,\\ Roy Maartens$^{d,e}$, Sabino Matarrese$^{a,b,f}$, Misao Sasaki$^{g}$, Licia Verde$^{h,i,l}$, David Wands$^{e}$\\~}

\affiliation{
$^a$Dipartimento di Fisica Galileo Galilei, Universit\`{a} di Padova,  I-35131 Padova, Italy\\
$^b$INFN Sezione di Padova,  I-35131 Padova, Italy\\
$^c$Argelander-Institut f\"ur Astronomie, D-53121 Bonn, Germany\\
$^d$Physics Department, University of the Western Cape, Cape Town 7535, South Africa\\
$^e$Institute of Cosmology \& Gravitation, University of Portsmouth, Portsmouth PO1 3FX, UK\\
$^f$Gran Sasso Science Institute, INFN, I-67100 L'Aquila, Italy\\
$^g$Yukawa Institute for Theoretical Physics, Kyoto University, Kyoto 606-8502, Japan\\
$^h$Instituci\'o Catalana de Recerca i Estudis Avan\c{c}at \& Instituto de ciencias del Cosmos, Universitat de Barcelona,  Barcelona 08028, Spain\\
$^i$ Radcliffe Institute for Advanced Study \& ITC, Centre for Astrophysics, Harvard University, MA 02138, USA\\ 
$^l$Institute of Theoretical Astrophysics, University of Oslo, Oslo 0315, Norway}

\begin{abstract}

In General Relativity, the constraint equation relating metric and density perturbations is inherently nonlinear, leading to an effective non-Gaussianity in the dark matter density field on large scales -- even if the primordial metric perturbation is Gaussian. 
Intrinsic non-Gaussianity in the large-scale dark matter overdensity in GR is real and physical. 
However, the variance smoothed on a local physical scale is not correlated with the large-scale curvature perturbation, 
so that there is no relativistic signature in the galaxy bias  when using the simplest model of bias. 
It is an open question whether the observable mass proxies such as luminosity or weak lensing correspond directly to the physical mass in the simple halo bias model. If not, there may be observables that encode this relativistic signature.

\end{abstract}

\date{\today}

\maketitle

\section{Introduction}

In Newtonian gravity, the Poisson equation is a linear relation between the gravitational potential and the matter overdensity. By contrast, in General Relativity (GR) this is replaced by a nonlinear relation, which introduces mode coupling between large and small scales~\cite{Bartolo:2005xa,Verde:2009hy,Bartolo:2010rw}. 
The original result has been confirmed by a number of independent calculations~\cite{Fitzpatrick:2009ci,Bruni:2013qta, Bruni:2014xma, Uggla:2013kya, Villa:2014foa,Villa:2015ppa}. 
Similar mode coupling can be produced 
in Newtonian gravity 
by local-type primordial non-Gaussianity of the gravitational potential \cite{Dalal:2007cu,Matarrese:2008nc}. 
The GR effect has therefore previously been interpreted as an effective local non-Gaussianity on very large scales~ { \cite{Bruni:2014xma,Carbone:2008iz,Carbone:2010sb,Camera:2014bwa, Camera:2014sba}.}

Recently two papers have argued that a ``separate universe'' approach can be used to show that {\em no} scale-dependent bias arises from the GR corrections on large scales \cite{Dai:2015jaa,dePutter:2015vga}. 
The {  argument} is that the nonlinear coupling between long-wavelength perturbations on a scale $\lambda_L$, and the small-scale variance, $\sigma_S^2=\langle\delta_S^2\rangle$, on a scale $\lambda_S$, vanishes under a local coordinate rescaling and hence is unobservable.

The separate universe approach \cite{Salopek:1990jq,Wands:2000dp} has proved to be a powerful tool to understand the origin of large-scale structure, and primordial non-Gaussianity, from inflation. Accelerated expansion in the very early Universe stretches initial small-scale vacuum fluctuations up to scales much larger than the Hubble scale at the end of inflation. Spatial gradients for such long-wavelength modes become small relative to the local Hubble time, and for many scales of interest, the perturbed universe can be treated as a patchwork of ``separate universes'', each locally obeying the classical {   Friedmann-Lema\^itre-Robertson-Walker (FLRW)} evolution of an unperturbed universe.

The separate universe approach is particularly  {  useful} for studying nonlinear perturbations on large scales \cite{Salopek:1990jq,Lyth:2005fi}. 
For adiabatic perturbations, each separate universe patch follows locally the same evolution as the unperturbed ``background'' cosmology.
The only difference between separate patches is the local expansion, characterised by the comoving metric perturbation $\zeta$. This is defined to be the local perturbation of the integrated expansion rate with respect to a background flat reference cosmology, $\delta N=N-\bar{N}$, where $N=\int dt\, \Theta/3$.

An important consequence of the uniqueness of the local evolution for adiabatic perturbations is that $\zeta$ is conserved on large scales where the separate universe approach is valid \cite{Wands:2000dp,Lyth:2004gb,Langlois:2005ii}. 
%
\begin{figure}[ht!]
  \centering{
  \includegraphics[width=8cm]{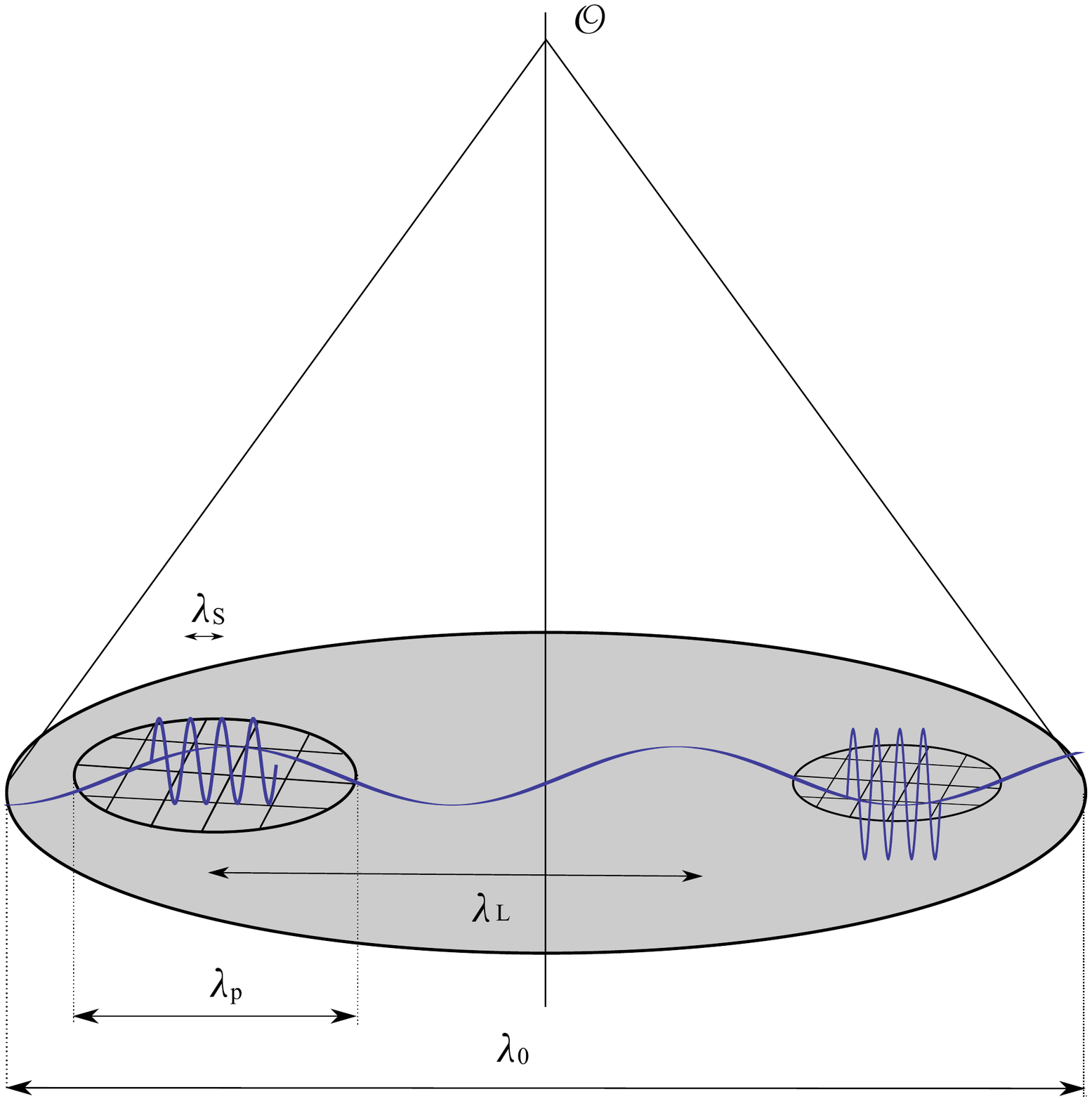}\vspace*{-2cm}
  }
  \caption{Schematic of the various scales in \eqref{scales}.  
   }
\end{figure}

In each patch,  the comoving spatial line element is (see \cite{dePutter:2015vga})
\be\label{met}
ds_{(3)}^2 = 
e^{2\zeta} \delta_{ij} dx^i dx^j \,.
\ee
There is a global {\em background} which must be defined with respect to some scale $\lambda_0$, at least as large as all the other scales of interest, i.e., at least as large as our presently observable Universe. It is important to distinguish this from the scale of the separate universe patches, $\lambda_P$. This is large enough for each patch to be treated as locally homogeneous and isotropic, but patches must be stitched together to describe the long-wavelength perturbations on a scale $\lambda_L \gg \lambda_P$. Thus, following \cite{Wands:2000dp}, we require a hierarchy of scales (see Fig.~1):
\be\label{scales}
\lambda_0 > \lambda_L \gg \lambda_P \gg \lambda_S\,.
\ee

The local observer in a separate universe patch cannot observe the effect of $\zeta_L$, which is locally homogeneous on the patch scale $\lambda_P$. However,  local coordinates can be defined only locally and the long mode curvature perturbation is observable through a mapping from local to global coordinates.

\section{The physical effect of curvature within the observable Universe}

In Newtonian gravity the only constraint on initial conditions is the Poisson equation, which provides a linear relation between the overdensity and the gravitational potential at all orders
\be
\label{Poisson}
\nabla^2 \Phi_N = - \frac{3}{2}a^2H^2 \delta \,.
\ee
Thus if the initial Newtonian potential $\Phi_N$ is Gaussian, then so is the initial density field $\delta$.
In GR, the nonlinear energy constraint equation for irrotational dust is~\cite{EMM_book}
\be
\frac23\Theta^2 - 2\sigma^2 + R^{(3)} = 16\pi G\rho + 2\Lambda \,,
\ee
where $\rho$ is the comoving matter density, $\Lambda$ is the cosmological constant, $\Theta=\nabla_\mu u^\mu$ is the expansion rate of the matter 4-velocity, $\sigma$ is its shear, and $R^{(3)}$ is the Ricci curvature scalar of the 3-dimensional space orthogonal to $u^\mu$.
At first order in perturbations about an FLRW cosmology, the energy constraint combines with the momentum constraint to give the relativistic version of the Poisson equation (\ref{Poisson}), where $\Phi_N$ is replaced by $\Phi$, i.e. the spatial metric perturbation in longitudinal gauge, and $\delta$ is the synchronous comoving gauge density contrast. Note that $\Phi=3\zeta/5$ at first order.
At second order, at the start of the matter era, using the relation  between $R^{(3)}$ and $\zeta$, we obtain \cite{Bruni:2014xma}
\be
\label{GR2}
\nabla^2 \zeta - 2\zeta\nabla^2\zeta + \frac12 \left(\nabla\zeta\right)^2  = - \frac{5}{2}a^2H^2 \delta  \,.
\ee

Consider a Gaussian distribution of $\zeta$. We separate $\zeta$ and  $\delta$ into independent long- and short-wavelength modes, $\zeta= \zeta_L + \zeta_S$ and $\delta= \delta_L + \delta_S$, where the wavelength of the long modes $\lambda_L$ obeys \eqref{scales}; in particular, $\lambda_L \gg \lambda_P$. 
To leading order in $\zeta_S$ and $\zeta_L$, and neglecting gradients of  $\zeta_L$ relative to those of $\zeta_S$, the initial constraint (\ref{GR2}) implies $ \nabla^2 \zeta_L=-5a^2 H^2\delta_L/2$ and
\begin{equation}
\nabla^2 \zeta_S - 2 \zeta_L \nabla^2 \zeta_S = - \frac{5 }{2} a^2 H^2\delta_S.
\label{longshort}
\end{equation}
The second term on the left represents the long-short mode coupling. 

Within a local patch on a scale $\lambda_P$, it is possible to redefine the background spatial coordinates to absorb the effects of the long-wavelength perturbations $\R_L$,
following \cite{dePutter:2015vga}:
\begin{equation}
\tilde{x}^i = x^i + \xi^i, \quad \xi^i = \R_L x^i.
\label{trans}
\end{equation}
If we neglect gradients of the long mode, this transformation eliminates $\R_L$ from the spatial metric \eqref{met}
\begin{equation}\label{zlmet}
ds_{(3)}^2 = e^{2 \R_L} e^{2 \R_S} \delta_{ij} dx^i dx^j 
= e^{2 \R_S} \delta_{ij} d \tilde{x}^i d \tilde{x}^j.  
\end{equation}
 This transformation holds for each {\em single} patch (see Fig.~1).

Since this is a purely spatial coordinate transformation, the curvature and density perturbations transform as scalars,
\begin{equation}\label{zdt}
\tilde{\R}_S(\tilde{x}) = \R_S(x)\,,~\tilde{\delta}_S(\tilde{x}) = \delta_S(x)\,,~~\mbox{where}~~ \tilde{x}=\big[1+\zeta_L(x)\big]x\,.
\end{equation}
The constraint equation (\ref{longshort}) becomes 
\begin{equation}
\tilde{\nabla}^2 \tilde{\R}_S(\tilde{x})  = - \frac{5 }{2}a^2 H^2 \tilde{\delta}_S (\tilde{x}).
\label{localPoisson1}
\end{equation}
Thus in the new local coordinates,  in one patch of size $\lambda_P$, we have a linear Poisson equation and the long-short mode coupling appears to be absent. 
This confirms the fact that the local observer in a separate universe patch cannot observe the effect of the locally homogeneous perturbation $\zeta_L$, as argued in~\cite{Dai:2015jaa,dePutter:2015vga}.

The original coordinates $x^i$ define a global chart, which is essential for defining random fields such as $\zeta_L$ on large scales, and the long mode curvature perturbation enters through the mapping from local to global coordinates. 
Indeed, a coordinate transformation that depends on a random field is not a
new concept in large-scale structure. The situation here is reminiscent of the redshift-space distortion map, where
 the random field is given by the peculiar velocities
(which are in turn generated by large-scale density perturbations). Because
of the nonlinear nature of this map, an initially Gaussian field in real space
becomes non-Gaussian in redshift space \cite{Scocc2004,Lewis2008}. 

As shown in \cite{dePutter:2015vga},   
$\tilde{\zeta}(\tilde{x})$ and $\tilde{\delta}(\tilde{x})$ are Gaussian fields with respect to the local coordinates $\tilde{x}^i$. By \eqref{zdt}, $\tilde{\delta} (\tilde{x})=\delta(x)$ and so  we recover a non-Gaussian distribution for the density field with respect to the global coordinates, $x^i$.
 See Fig.~2 for a  schematic illustration of this. 
 %
\begin{figure}[ht!]
  \centering{
  \includegraphics[width=10cm]{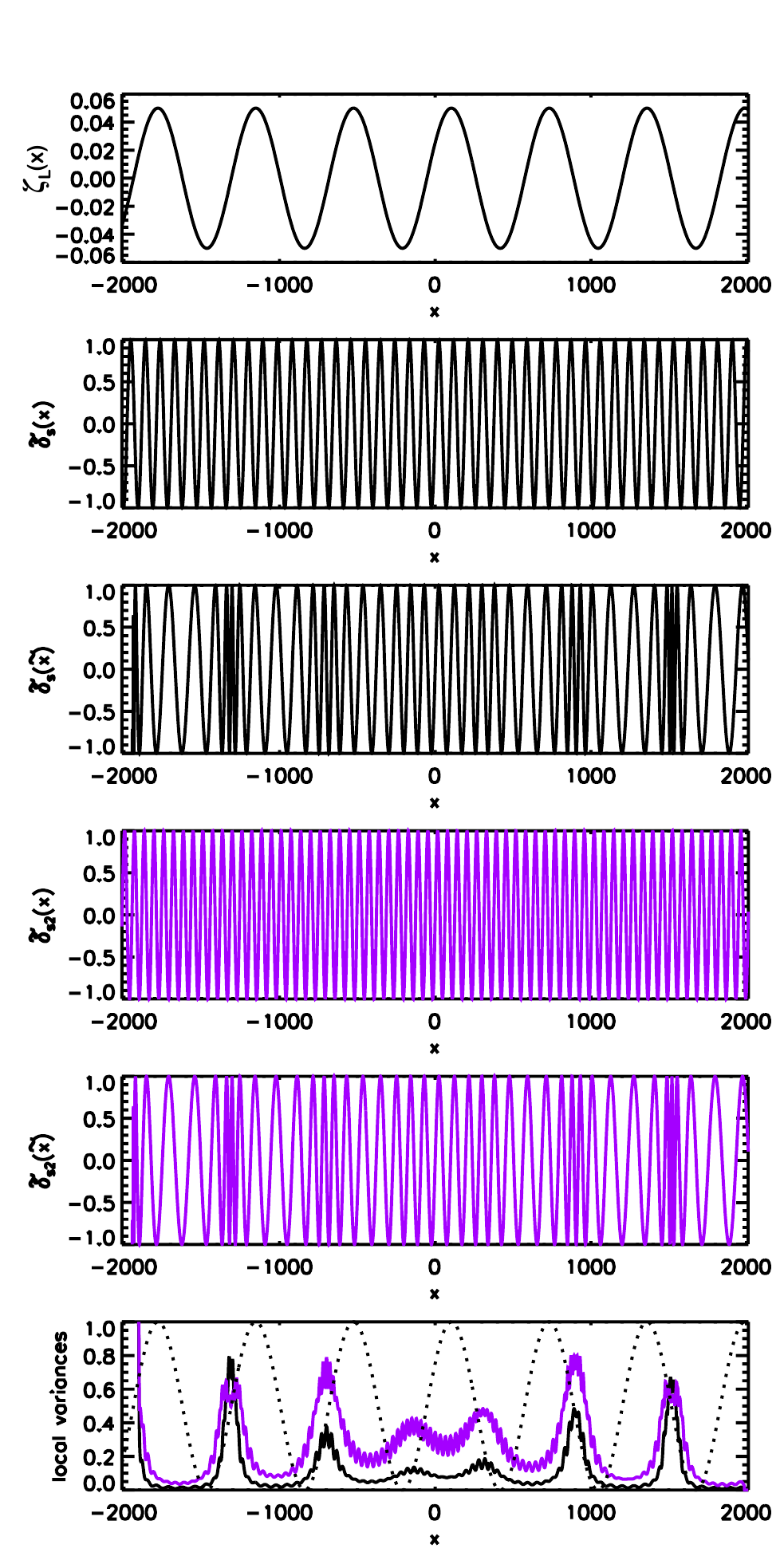}
  }
  \caption{ Illustrating the non-Gaussianity of $\tilde{\delta}_S(\tilde{x})$ and its correlation with $\zeta_L$ in the global coordinates, $x$. 
  We start from the fact that $\tilde{\delta}_S(x)$ is Gaussian \cite{dePutter:2015vga}, and uncorrelated with the Gaussian  $\zeta_L(x)$. The top 2 panels show one  $k$ mode for each.
Then we apply the transformation \eqref{zdt}, $x\to \tilde{x}=x[1+\zeta_L(x)]$.  The resulting $\tilde{\delta}_S (\tilde{x})$ field  (next panel down) is clearly modulated by $\zeta_L$. To see that it is  non-Gaussian,  consider another $k$ mode, $\tilde{\delta}_{S2}(x)$ (fourth panel). Clearly $\tilde{\delta}_{S2}(\tilde{x})$  is highly correlated to  $\tilde{\delta}_S(\tilde{x})$, i.e., there are phase correlations.
The local variances of these two modes at fixed coordinate scale are correlated with $\zeta_L$ (bottom panel).\\
   }
\end{figure}
  
There is no argument against the fact that $\tilde{x}^i$ coordinates are useful to discuss physics in a local patch of size $\lambda_P \ll \lambda_L$ --
but this applies only locally. The effect of the long mode $\R_L$ is to create spatial curvature $\nabla^2 \R_L$ on a constant-time hypersurface. The curvature can be eliminated only locally, by neglecting gradients of $\zeta_L$, as in \eqref{zlmet}. Beyond the single patch, when gradients are not negligible, we have \cite{Manasse:1963zz}
\begin{equation}
ds_{(3)}^2 = e^{2 \R_L} e^{2 \R_S} \delta_{ij} dx^i dx^j 
= e^{2 \R_S} \Big[\delta_{ij} d \tilde{x}^i d \tilde{x}^j + O\big(|\tilde{x}|^2 \nabla^2 \R_L \big) \Big] \,.
\end{equation}

Indeed, all coordinate transformations that neglect curvature can only be defined locally (see \cite{Manasse:1963zz}).
The spatial curvature generated by $\R_L$ is directly related to the long-wavelength density perturbation $\delta_L$ by the long-wavelength part of \eqref{GR2}. On a constant-time hypersurface, the density perturbation cannot be eliminated by the spatial coordinate transformation. This implies that we need many different local patches of scale $\sim\lambda_P$ described by different $\tilde{x}^i$-coordinates in the entire observed universe of scale $\sim\lambda_0$. 

Within a single patch on a scale $\lambda_P$, the local observer does not notice this stochastic nature of the local coordinates. However, once we are interested in physics beyond the local patch  we have to compare the different patches and we notice that $\tilde{x}^i$ vary stochastically through their dependence on $\R_L$. 
This implies that GR non-Gaussianity is present in the dark matter distribution.  
Nonlinear GR effects   can be appreciated only by looking at a region where  $\nabla^2  \zeta_L$ is not negligible and, in principle, the stochastic nature of $\zeta_L$ could become apparent.

\section{Local-type non-Gaussianity in the Newtonian density field}

In Newtonian gravity, it is standard to parametrise primordial non-Gaussianity of local type in the Newtonian potential by 
\begin{equation}
\Phi_N = \varphi + f_{\rm NL} \left( \varphi^2 - \langle \varphi^2 \rangle \right),
\end{equation}
where $\varphi$ is a Gaussian random field.
If we split the Gaussian field into long and short modes, $\varphi=\varphi_S+\varphi_L$, and use the same assumptions that lead to \eqref{longshort}, then the Poisson equation (\ref{Poisson}) yields
\be
\label{fnlPoisson}
\big( 1+2 f_{\rm NL} \varphi_L \big) \nabla^2 \varphi_S = - \frac{3}{2} a^2H^2\delta_S \,.
\ee
This leads to a scale-dependent galaxy bias~\cite{Dalal:2007cu,Matarrese:2008nc}.

Single-field, slow-roll inflation generates an almost Gaussian distribution for $\zeta$, which remains Gaussian for adiabatic perturbations on super-Hubble scales through to the start of the matter-dominated era.
For this Gaussian case, using the first-order relation $\zeta=5\varphi/3$, the GR second-order  constraint equation (\ref{longshort}) gives
\begin{eqnarray}
\Big( 1-{10\over 3} \varphi_L \Big) \nabla^2 \varphi_S = - \frac{3}{2} a^2H^2\delta_S. \label{fgr}
 \end{eqnarray} 
This has been used to argue that GR corrections to the comoving density field will also lead to a scale-dependent bias, with an  effective (local) non-Gaussianity parameter~{ \cite{Bruni:2014xma, Carbone:2008iz,Carbone:2010sb,Camera:2014bwa, Camera:2014sba}:}
\be\label{fgr2}
f_{\rm NL}^{\rm GR,eff}=-{5\over 3}\,.
\ee

While the Newtonian potential $\Phi_N$ satisfies the linear Poisson equation at all orders, in GR $\zeta$ obeys the nonlinear constraint (\ref{GR2}).  We could thus define primordial non-Gaussianity of the comoving density field, e.g., from inflation, in terms of the primordial metric perturbation, $\zeta$. Local-type primordial non-Gaussianity can be parametrised as
\be
{  \zeta = \frac{5}{3} \Big[ 
 \varphi + 
  f_{\rm NL}
  \left( \varphi^2 - \langle \varphi^2 \rangle \right) \Big]. 
}
\ee 
Thus,
the non-Gaussianity present in the comoving density field (\ref{GR2}) is similar to a Newtonian density field with local-type non-Gaussianity parameter
\be
{  f^{\rm eff}_{\rm NL}
=f_{\rm NL}-{5\over3}\,.
}
\ee
However, as we shall show in the next section, this Newtonian description of the density field in terms of an effective $\fnl$ is {\em not} sufficient to describe local halo abundance in GR.
{  When considering halos, smoothing of the dark matter field must be introduced.  
 The variance of $\tilde{\delta}(\tilde{x})$ smoothed on a physical scale is not correlated with the large-scale modes $\zeta_L$, unlike the case with primordial non-Gaussianity of local type. 
}

\section{Scale-dependent bias and single-field inflation}

The presence of a large-scale perturbation can change the local abundance of collapsed dark matter halos either through a perturbation in the local ``background" density, $\delta_L$, or through a perturbation in the local variance, $\sigma_R\to\bar\sigma_R+\delta\sigma_R$, on a comoving scale $R\ll \lambda_P$, where we use a bar to denote quantities in the global background.

For a Gaussian density field the small-scale variance is independent of the long-wavelength perturbations, 
but for a non-Gaussian field of the local-type in Newtonian gravity, as given in (\ref{fnlPoisson}), the small-scale variance is modulated by long-wavelength perturbations and we have
\be
\sigma_{\bar R}^2\big|_{\varphi_L} = (1+4f_{\rm NL}\varphi_L) \bar\sigma_{\bar R}^2 \,.
\ee
Here 
$\sigma_{\bar{R}}^2$ is the variance of the density field smoothed on the comoving length scale $\bar{R}$,
corresponding to a given physical mass in the global background,
\be
 \label{def:barM}
M = \frac43 \pi \bar{R}^3 \bar\rho\,.
\ee
We have $\delta\sigma_R\propto \fnl\varphi_L\propto \fnl\delta_L/k_L^2$ and hence local-type non-Gaussianity can lead to a strongly scale-dependent bias for halos~\cite{Dalal:2007cu,Matarrese:2008nc}.

In GR the long-wavelength metric perturbation, $\zeta_L$, also leads to a change in the variance at a fixed comoving radius, $R$. The density perturbations obey the nonlinear constraint equation. This nonlinearity modulates the density field by $\zeta_L$:  
\begin{equation}
\delta(x)\big|_{\zeta_L} = (1 - 2 \zeta_L) \delta(x),
\end{equation}
where $\delta(x)$ refers to the density perturbations in the absence of $\zeta_L$ while $|_{\zeta_L}$ denotes that in the presence of $\zeta_L$. From (\ref{longshort}) we have
\be
 \label{sigRgivenzeta}
\sigma_R^2\big|_{\zeta_L} = (1-4\zeta_L) \bar\sigma_R^2 \,.
\ee
However in GR we must also account for the perturbation in the mass enclosed within a comoving radius $R$ due to a long-wavelength metric perturbation $\zeta_L$:
\be
\label{pertM}
M = \frac43 \pi (1+3\zeta_L) R^3 \rho \,.
\ee
To compare the local abundance of halos of a fixed physical mass, $M$ in (\ref{def:barM}), in the presence of a long-wavelength perturbation, $\zeta_L$, we must compare properties of the density field on a fixed local {\em physical} scale $(1+\zeta_L)R=\bar{R}$. Hence
\be
 \label{RvsbarR}
R = (1-\zeta_L)\bar{R} \,.
\ee
Note that we neglect the perturbation in the density, $\rho$, in (\ref{pertM}), since $\delta_L$ is negligible relative to $\varphi_L$ for long-wavelength perturbations. 

For a scale-invariant distribution of the curvature perturbation, $\zeta$, the variance of the density field at two different comoving scales $R_1$ and $R_2$ are related by
\be
\bar\sigma_{R_1}^2 = \left( \frac{R_1}{R_2} \right)^{-4} \bar\sigma_{R_2}^2 \,,
\ee
for scales greater than the matter-radiation equality horizon scale, $R_1,R_2>\lambda_{\rm eq}$.
Using (\ref{RvsbarR}), we obtain
\be
 \label{sigRvsbarR}
\bar\sigma_{R}^2 = \left( 1+4\zeta_L \right) \bar\sigma_{\bar{R}}^2 \,.
\ee

Thus combining (\ref{sigRgivenzeta}) and~(\ref{sigRvsbarR}), we obtain
\be
\sigma_R^2\big|_{\zeta_L} = \bar\sigma_{\bar{R}}^2 \,.
\ee

As shown in (\ref{localPoisson1}), the small-scale density at a fixed {\em local  physical} scale is independent of the long-wavelength perturbation. Thus the long-wavelength mode has no effect on the small-scale variance of the density field smoothed on a fixed mass scale.

Although there is a change in the local variance at a fixed comoving coordinate radius, $R$, this is exactly compensated by the change in the local coordinate scale corresponding to a fixed physical mass and the scale-dependence of the variance \cite{dePutter:2015vga}.
This is consistent with the argument that the long-wavelength perturbations $\lambda_L\gg\lambda_P$ are not observable locally if the distribution of the primordial metric perturbation, $\zeta$, is a Gaussian random field \cite{Dai:2015jaa,dePutter:2015vga}.

Note that in Newtonian gravity there is no equivalent perturbation in the physical volume and hence there is no perturbation in the mass enclosed at a coordinate radius $R$ caused by the large-scale potential $\varphi_L$. 
{  
A Newtonian description with $f_{\rm NL}^{\rm GR,eff}=-5/3$ thus fails to account for the local coordinate invariance inherent in GR.}

\section{Conclusions}

We have shown how non-Gaussian correlations in the matter overdensity arise due to nonlinear constraints in GR, even when the primordial metric perturbation from inflation, $\R$, is described by a Gaussian random field. 
This may be understood in a simple way as arising from the long-wavelength metric perturbation $\R_L$ rescaling the local small-scale curvature perturbation $\R_S$, and thus the local small-scale density field, through the nonlinear constraint equation (\ref{longshort}).

A very long wavelength metric perturbation can be removed locally by a coordinate transformation \eqref{trans}.
If the long-wavelength perturbation were much larger than our observable horizon, $\lambda_L\gg\lambda_0$, then that would be the end of the story --  {  perturbations with wavelength} much larger than our horizon form part of our background cosmology and cannot be observed locally.
However, local coordinates can be defined only locally and the long mode curvature perturbation enters through a mapping from local to global coordinates.
{  This coordinate transformation depends on a random field and is reminiscent of the mapping between real and redshift space, where the large-scale random field is given by the velocity field. It is known that  redshift space distortions  induce non-Gaussianity}.

The fact that the nonlinear GR effect is present in global coordinates,  is similar to  
 non-Gaussianities in the CMB that arise at recombination due solely to the intrinsic nonlinearity of relativistic perturbations~\cite{Mollerach,Bartolo:2004ty,Creminelli:2004pv}. If the long-wavelength perturbations are on scales greater than the present horizon, they cannot have any physical effect, and can be rescaled away (this corresponds to properly redefining the background average temperature~\cite{Boubekeur:2009uk,Creminelli:2011sq}). However, we are interested in  modes that are inside the horizon today, and we need to compare different patches of the sky modulated by a long mode \cite{Creminelli:2011sq,Mirbabayi:2014hda}. This provides an alternative understanding of a GR term that was missing in~\cite{Creminelli:2004pv}, compared to the expression for the squeezed CMB angular bispectrum obtained in~\cite{Bartolo:2004ty,Bartolo:2011wb,Creminelli:2011sq}.
 The angular bispectrum of CMB temperature anisotropies (and polarisation) has been estimated~\cite{Bartolo:2004ty,Creminelli:2011sq} and shown to reproduce in this squeezed limit the full numerical calculations of the Einstein-Boltzmann system at second order~\cite{Huang:2012ub,Su:2012gt,Pettinari:2013he,Pettinari:2014iha}. This intrinsic non-Gaussianity in the CMB, predicted by local rescaling arguments, could in principle be observed by future experiments.

On the other hand, to {compute} the local abundance of halos of a fixed physical mass, we have to introduce smoothing on a local physical scale $\tilde{R}=(1+\zeta_L)R$ and the corresponding local variance is independent of $\zeta_L$, {  as argued by~\cite{dePutter:2015vga}.}
Thus the short-wavelength variance {\em at a fixed physical mass scale} is not modulated by the long-wavelength metric perturbation in GR.

{   It is still an open question whether observable mass proxies such as luminosity,  velocity dispersion, X-ray emission, SZ decrement, or gravitational weak lensing signal,  correspond directly to the fixed physical mass defined above.} 
Intrinsic non-Gaussianity in the large-scale density field in GR is real and physical. It could leave observable signatures in the large-scale structure of the Universe,  provided we can find observable quantities  that depend on global  variables  such as the distance between the observer and a distant patch.

\[\]
{\bf Acknowledgments:} 
We thank  {  Phil Bull}, {  Chris Clarkson}, Liang Dai,  {Roland de Putter}, {Olivier Dor\'e}, Daniel Green, {Alan Heavens}, {  Anthony Lewis}, Enrico Pajer, Alvise Raccanelli, Cornelius Rampf,   {Antonio Riotto}, Fabian Schmidt,  {Martin Sloth}, Obinna Umeh,  Eleonora Villa, {   Matthias Zaldarriaga} for useful discussions.
MB, KK, MS and DW benefited from discussions during the workshop {\em Relativistic Cosmology} (YITP-T-14-04) at the Yukawa Institute for Theoretical Physics, Kyoto University. 
NB and SM acknowledge partial financial support from the ASI/INAF Agreement 2014-024-R.0 for the Planck LFI Activity of Phase E2. 
DB is supported by the Deutsche Forschungsgemeinschaft through Transregio 33, {\em The Dark Universe}. 
RM is supported by the South African Square Kilometre Array Project and the South African National Research Foundation. MB, KK, RM, and DW are supported by the UK Science \& Technology Facilities Council grant ST/K00090X/1. MB, KK, RM and DW are also supported by STFC grant ST/L005573/1.  KK is supported by the European Research Council grant through 646702 (CosTesGrav). 
MS is supported by JSPS Grant-in-Aid for Scientific Research (A) No.~21244033. LV acknowledges support   from the European Research Council (grant FP7-IDEAS-Phys.LSS 240117)  a Mineco grant 
FPA2011-29678-C02-02, and  MDM-2014-0369 of ICCUB (Unidad de Excelencia ``Mar\'ia de Maeztu'').


\begin{thebibliography}{99}


\bibitem{Bartolo:2005xa}
  N.~Bartolo, S.~Matarrese and A.~Riotto,
  ``Signatures of primordial non-Gaussianity in the large-scale structure of the Universe,''
  JCAP {\bf 0510}, 010 (2005)
  [astro-ph/0501614].


\bibitem{Verde:2009hy}
  L.~Verde and S.~Matarrese,
  ``Detectability of the effect of Inflationary non-Gaussianity on halo bias,''
  Astrophys.\ J.\  {\bf 706}, L91 (2009)
  [arXiv:0909.3224].
  
\bibitem{Bartolo:2010rw}
  N.~Bartolo, S.~Matarrese, O.~Pantano and A.~Riotto,
  ``Second-order matter perturbations in a $\Lambda$CDM cosmology and non-Gaussianity,''
  Class.\ Quant.\ Grav.\  {\bf 27}, 124009 (2010)
  [arXiv:1002.3759].

\bibitem{Fitzpatrick:2009ci}
  A.~L.~Fitzpatrick, L.~Senatore and M.~Zaldarriaga,
  ``Contributions to the Dark Matter 3-Pt Function from the Radiation Era,''
  JCAP {\bf 1005}, 004 (2010)
  [arXiv:0902.2814].

\bibitem{Bruni:2013qta}
  M.~Bruni, J.~C.~Hidalgo, N.~Meures and D.~Wands,
  ``Non-Gaussian Initial Conditions in $\Lambda$CDM: Newtonian, Relativistic, and Primordial Contributions,''
  Astrophys.\ J.\  {\bf 785}, 2 (2014)
  [arXiv:1307.1478].
  
\bibitem{Bruni:2014xma}
  M.~Bruni, J.~C.~Hidalgo and D.~Wands,
  ``Einstein's signature in cosmological large-scale structure,''
  Astrophys.\ J.\  {\bf 794},  L11 (2014)
  [arXiv:1405.7006].


\bibitem{Uggla:2013kya}
  C.~Uggla and J.~Wainwright,
  ``Simple expressions for second order density perturbations in standard cosmology,''
  Class.\ Quant.\ Grav.\  {\bf 31}, 105008 (2014)
  [arXiv:1312.1929].                  
                                             
\bibitem{Villa:2014foa}
  E.~Villa, L.~Verde and S.~Matarrese,
  ``General relativistic corrections and non-Gaussianity in large-scale structure,''
  Class.\ Quantum\ Grav.\ {\bf 31}, 234005 (2014)
  [arXiv:1409.4738].

\bibitem{Villa:2015ppa} 
  E.~Villa and C.~Rampf,
  ``Relativistic perturbations in $\Lambda$CDM: Eulerian and Lagrangian approaches,''
  arXiv:1505.04782.
  
\bibitem{Dalal:2007cu}
  N.~Dalal, O.~Dor\'e, D.~Huterer and A.~Shirokov,
  ``The imprints of primordial non-gaussianities on large-scale structure: scale dependent bias and abundance of virialized objects,''
  Phys.\ Rev.\ D {\bf 77}, 123514  (2008)
  [arXiv:0710.4560].
  
\bibitem{Matarrese:2008nc}
  S.~Matarrese and L.~Verde,
  ``The effect of primordial non-Gaussianity on halo bias,''
  Astrophys.\ J.\  {\bf 677},  L77 (2008)
  [arXiv:0801.4826].
  
 
  \bibitem{Carbone:2008iz} 
  C.~Carbone, L.~Verde and S.~Matarrese,
  ``Non-Gaussian halo bias and future galaxy surveys,''
  Astrophys.\ J.\  {\bf 684}, L1 (2008)
  [arXiv:0806.1950].
    
  \bibitem{Carbone:2010sb} 
  C.~Carbone, O.~Mena and L.~Verde,
  ``Cosmological Parameters Degeneracies and Non-Gaussian Halo Bias,''
  JCAP {\bf 1007}, 020 (2010)
  [arXiv:1003.0456].
  
\bibitem{Camera:2014bwa}
  S.~Camera, M.~G.~Santos and R.~Maartens,
  ``Probing primordial non-Gaussianity with SKA galaxy redshift surveys: a fully relativistic analysis,''
  Mon.\ Not.\ Roy.\ Astron.\ Soc.\  {\bf 448},  1035 (2015)
  [arXiv:1409.8286].

\bibitem{Camera:2014sba} 
S.~Camera, R.~Maartens and M.~G.~Santos,
  ``Einstein's legacy in galaxy surveys,'' Mon.\ Not.\ Roy.\ Astron.\ Soc.\ Lett., in press (2015)
  [arXiv:1412.4781].
  
\bibitem{Dai:2015jaa}
  L.~Dai, E.~Pajer and F.~Schmidt,
  ``On Separate Universes,''
  arXiv:1504.00351.

\bibitem{dePutter:2015vga}
  R.~de Putter, O.~Dor\'e and D.~Green,
  ``Is There Scale-Dependent Bias in Single-Field Inflation?,''
  arXiv:1504.05935.

\bibitem{Salopek:1990jq} 
  D.~S.~Salopek and J.~R.~Bond,
  ``Nonlinear evolution of long wavelength metric fluctuations in inflationary models,''
  Phys.\ Rev.\ D {\bf 42}, 3936 (1990).

  
\bibitem{Wands:2000dp}
  D.~Wands, K.~A.~Malik, D.~H.~Lyth and A.~R.~Liddle,
  ``A New approach to the evolution of cosmological perturbations on large scales,''
  Phys.\ Rev.\ D {\bf 62}, 043527 (2000)
  [astro-ph/0003278].
         
\bibitem{Lyth:2005fi} 
  D.~H.~Lyth and Y.~Rodriguez,
  ``The Inflationary prediction for primordial non-Gaussianity,''
  Phys.\ Rev.\ Lett.\  {\bf 95}, 121302 (2005)
  [astro-ph/0504045].

\bibitem{Lyth:2004gb} 
  D.~H.~Lyth, K.~A.~Malik and M.~Sasaki,
  ``A General proof of the conservation of the curvature perturbation,''
  JCAP {\bf 0505}, 004 (2005)
  [astro-ph/0411220].
  
\bibitem{Langlois:2005ii} 
  D.~Langlois and F.~Vernizzi,
  ``Evolution of nonlinear cosmological perturbations,''
  Phys.\ Rev.\ Lett.\  {\bf 95}, 091303 (2005)
  [astro-ph/0503416].
  
\bibitem{EMM_book}
   G.~F.~R.~Ellis, R.~Maartens and M.~A.~H.~MacCallum,
 {\it Relativistic Cosmology}, Cambridge University Press (2012).
         
\bibitem{Scocc2004}
   R.~Scoccimarro,
  ``Redshift-space distortions, pairwise velocities and nonlinearities,''
  Phys.\ Rev.\ D {\bf 70}, 083007 (2004)
  [astro-ph/0407214].
 
    \bibitem{Lewis2008}
 J.~R.~Shaw and A.~Lewis,
  ``Nonlinear Redshift-Space Power Spectra,''
  Phys.\ Rev.\ D {\bf 78}, 103512 (2008)
  [arXiv:0808.1724].

\bibitem{Manasse:1963zz} 
  F.~K.~Manasse and C.~W.~Misner,
  ``Fermi Normal Coordinates and Some Basic Concepts in Differential Geometry,''
  J.\ Math.\ Phys.\  {\bf 4}, 735 (1963).
       

\bibitem{Mollerach}
S.~Mollerach and S.~Matarrese,
``Cosmic microwave background anisotropies from second order gravitational perturbations,''
Phys.\ Rev.\ D {\bf 56}, 4494 (1997)
[astro-ph/9702234].

\bibitem{Bartolo:2004ty} 
  N.~Bartolo, S.~Matarrese and A.~Riotto,
  ``Gauge-invariant temperature anisotropies and primordial non-Gaussianity,''
  Phys.\ Rev.\ Lett.\  {\bf 93}, 231301 (2004)
  [astro-ph/0407505].
  
\bibitem{Creminelli:2004pv} 
  P.~Creminelli and M.~Zaldarriaga,
  ``CMB 3-point functions generated by nonlinearities at recombination,''
  Phys.\ Rev.\ D {\bf 70}, 083532 (2004)
  [astro-ph/0405428].

\bibitem{Creminelli:2011sq} 
  P.~Creminelli, C.~Pitrou and F.~Vernizzi,
  ``The CMB bispectrum in the squeezed limit,''
  JCAP {\bf 1111}, 025 (2011)
  [arXiv:1109.1822].

\bibitem{Boubekeur:2009uk} 
  L.~Boubekeur, P.~Creminelli, G.~D'Amico, J.~Norena and F.~Vernizzi,
  ``Sachs-Wolfe at second order: the CMB bispectrum on large angular scales,''
  JCAP {\bf 0908}, 029 (2009)
  [arXiv:0906.0980].


\bibitem{Mirbabayi:2014hda}
  M.~Mirbabayi and M.~Zaldarriaga,
  ``CMB Anisotropies from a Gradient Mode,''
  JCAP {\bf 1503}, 056 (2015)
  [arXiv:1409.4777].

 \bibitem{Bartolo:2011wb} 
  N.~Bartolo, S.~Matarrese and A.~Riotto,
  ``Non-Gaussianity in the Cosmic Microwave Background Anisotropies at Recombination in the Squeezed limit,''
  JCAP {\bf 1202}, 017 (2012)
  [arXiv:1109.2043].

  
\bibitem{Huang:2012ub} 
  Z.~Huang and F.~Vernizzi,
  ``Cosmic Microwave Background Bispectrum from Recombination,''
  Phys.\ Rev.\ Lett.\  {\bf 110},  101303 (2013)
  [arXiv:1212.3573].
  
\bibitem{Su:2012gt} 
  S.-C.~Su, E.~A.~Lim and E.~P.~S.~Shellard,
 ``Cosmic microwave background bispectrum from nonlinear effects during recombination,''
  Phys.\ Rev.\ D {\bf 90}, 023004 (2014)
  [arXiv:1212.6968].
  
\bibitem{Pettinari:2013he} 
  G.~W.~Pettinari, C.~Fidler, R.~Crittenden, K.~Koyama and D.~Wands,
  ``The intrinsic bispectrum of the Cosmic Microwave Background,''
  JCAP {\bf 1304}, 003 (2013)
  [arXiv:1302.0832].
  
\bibitem{Pettinari:2014iha} 
  G.~W.~Pettinari, C.~Fidler, R.~Crittenden, K.~Koyama, A.~Lewis and D.~Wands,
  ``Impact of polarization on the intrinsic cosmic microwave background bispectrum,''
  Phys.\ Rev.\ D {\bf 90}, 103010 (2014)
  [arXiv:1406.2981].
    


\end{thebibliography}
\end{document}